\documentstyle[11pt]{article}
\newlength{\bredde}
\def\slash#1{\settowidth{\bredde}{$#1$}\ifmmode\,\raisebox{.15ex}{/}
\hspace*{-\bredde} #1\else$\,\raisebox{.15ex}{/}\hspace*{-\bredde} #1$\fi}
\newcommand{\mat}{\left ( \begin{array}{cc}}
\newcommand{\emat}{\end{array} \right )}
\newcommand{\matt}{\left ( \begin{array}{ccc}}
\newcommand{\ematt}{\end{array} \right )}
\newcommand{\matf}{\left ( \begin{array}{cccc}}
\newcommand{\ematf}{\end{array} \right )}
\newcommand{\vect}{\left ( \begin{array}{c}}
\newcommand{\evect}{\end{array} \right )}

\newcommand{\be}{\begin{eqnarray}}
\newcommand{\ee}{\end{eqnarray}}
\newcommand{\beq}{\begin{equation}}
\newcommand{\eeq}{\end{equation}}
\newcommand{\ba}{\begin{array}{ccc}}
\newcommand{\ea}{\end{array}}

\newcommand{\noi}{\vspace{12pt}\noindent}
\newcommand{\lG}{\raise.3ex\hbox{$\stackrel{\leftarrow}{G}$}}
\newcommand{\lU}{\raise.3ex\hbox{$\stackrel{\leftarrow}{U}$}}
\newcommand{\lP}{\raise.3ex\hbox{$\stackrel{\leftarrow}{{\cal P}}$}}
\newcommand{\leta}{\raise.3ex\hbox{$\stackrel{\leftarrow}{\eta}$}}
\newcommand{\lOmega}{\raise.3ex\hbox{$\stackrel{\leftarrow}{\Omega}$}}
\newcommand{\ldr}{\raise.3ex\hbox{$\stackrel{\leftarrow}{\delta^r}$}}

\def\beqn{\begin{eqnarray}}
\def\eeqn{\end{eqnarray}}

\def\gtwid{\raise.3ex\hbox{$>$\kern-.75em\lower1ex\hbox{$\sim$}}}
\def\ltwid{\raise.3ex\hbox{$<$\kern-.75em\lower1ex\hbox{$\sim$}}}

\def\t{{\mbox{\rm Tr}}}
\begin{document}
\topmargin -1.4cm
\oddsidemargin -0.8cm
\evensidemargin -0.8cm
\title{\Large{{\bf Wilson Loops in ${\cal N}=4$ Supersymmetric
Yang-Mills Theory\\ from Random Matrix Theory}}}

\vspace{1.5cm}

\author{~\\{\sc G. Akemann}\\Max-Planck-Institut f\"ur Kernphysik\\
Saupfercheckweg 1\\D-69117 Heidelberg\\Germany\\~\\and\\~\\
{\sc P.H. Damgaard}\footnote{On leave from:
The Niels Bohr Institute, Blegdamsvej 17, DK-2100 Copenhagen,
Denmark.}\\Theory Division\\CERN\\CH-1211 Geneva 23\\Switzerland
}
\date{}
\maketitle
\vfill
\begin{abstract}
Based on the AdS/CFT correspondence, string theory has given exact predictions
for circular Wilson loops in $U(N)$ ${\cal N}=4$
supersymmetric Yang-Mills theory to all orders in a $1/N$ expansion.
These Wilson loops can also be derived from Random Matrix Theory.
In this paper we show that the result is generically insensitive to
details of the Random Matrix Theory potential. We also compute all
higher $k$-point correlation functions, which are needed for the evaluation
of Wilson loops in arbitrary irreducible representations of $U(N)$.
\end{abstract}
\vfill

\begin{flushleft}
CERN-TH/2001-026\\
hep-th/0101225
\end{flushleft}
\thispagestyle{empty}
\newpage

\section{Introduction}
One of the surprising predictions of the Maldacena conjecture \cite{M}
relating type IIB string theory on an $AdS_5\times S^5$ background to
${\cal N}=4$ supersymmetric Yang-Mills (SYM) theory in four dimensions
concerns the static quark-antiquark potential.
At strong `t Hooft gauge coupling
$\lambda \equiv g^2N$ this potential is predicted to be of strength
$\sqrt{\lambda}$ \cite{RY,M1} rather than $\lambda$ (as at weak coupling).
Because the gauge coupling does not run in ${\cal N}=4$ SYM this could
in principle be tested by explicit gauge theory computations.
Substantial progress in this direction was made last year by
Erickson, Semenoff and Zarembo \cite{ESZ}. They computed the sum
of all Feynman-gauge planar diagrams without internal vertices to
the expectation value of a circular Wilson loop in ${\cal N}=4$ SYM
and found ($I_n(x)$ is the $n$th order modified Bessel function)
\beq
\langle W \rangle_{circle} ~=~ 
\frac{2}{\sqrt{\lambda}}I_1(\sqrt{\lambda}) ~\sim~
\sqrt{\frac{2}{\pi}}\frac{e^{\sqrt{\lambda}}}{\lambda^{3/4}} ~,\label{Wesz}
\eeq
in agreement with the leading-order (in $\lambda$)
AdS/CFT prediction \cite{BCFM,DGO}
\beq
\langle W \rangle_{AdS/CFT} ~=~  e^{\sqrt{\lambda}} ~.
\eeq
By an explicit computation it was also shown in ref. \cite{ESZ} that
the first corrections to (\ref{Wesz}) of order $\lambda^2$ coming from
diagrams with internal vertices precisely cancel in four dimensions.

Recently, Drukker and Gross \cite{DG,G} have made a quite remarkable extension
of this result by (a) outlining a proof that the above sum of
rainbow diagrams (\ref{Wesz}) actually gives the exact result to all orders
in $\sqrt{\lambda}$ at $N = \infty$, and (b) showing that the
calculation can be extended to all orders in a $1/N^2$ expansion.
Comparing again with the AdS/CFT correspondence they find {\em exact} agreement
to leading order in $\lambda$, at every order in $1/N^2$.

One of the key ideas of ref. \cite{DG} is to make efficient use of
a Random Matrix Theory (RMT) interpretation \cite{ESZ} of the result
(\ref{Wesz}). Consider the unitary ensemble of $N\times N$ hermitian
matrices, and a corresponding partition function ${\cal Z}$ of Gaussian
Boltzmann factor. Identifying $N$ with that of the gauge group $U(N)$ of
${\cal N}=4$ SYM, it was noticed in \cite{ESZ} that in the limit
$N \to \infty$
\be
\left\langle \frac{1}{N}\t ~e^{M}\right\rangle &\equiv&
\frac{1}{{\cal Z}}\int\! dM~ \frac{1}{N} \t~e^{M}~
\exp\left[-\frac{2N}{\lambda}\t ~M^2\right] \cr
&=& \frac{2}{\sqrt{\lambda}}I_1(\sqrt{\lambda}) ~,
\ee
precisely reproduces the gauge theory result (\ref{Wesz}). Intuitive
reasons for why this is a correct representation of the circular
Wilson loop in ${\cal N}=4$ SYM have been given in refs. \cite{ESZ,DG}.
Moreover, in ref. \cite{DG} it was argued that this particular
Random Matrix Theory representation
can be used to compute exactly the leading order (in $\sqrt{\lambda}$)
circular Wilson loop to {\em all orders} in the $1/N^2$ expansion. It
should thus be possible to completely by-pass the otherwise quite
cumbersome computation in the ${\cal N}=4$ supersymmetric
gauge theory language.

The fact that the RMT representation was chosen with Gaussian Boltzmann
weight is clearly linked directly to the fact that only rainbow graphs
are being summed in the field theory context. An obvious question to
ask is why this can yield the full answer. Although arguments have been
given \cite{ESZ,DG}, a full proof is still lacking. While
the resulting effective theory of the circular Wilson loop can be argued
to be zero-dimensional (basically on account of a ``conformal
anomaly'' invalidating the conformal mapping from the line to the circle
at just one space-time point \cite{DG}), there is {\em a priori}
no guarantee that
vertices cannot contribute at that particular point. Such interaction
vertices would correspond to higher order terms in a more general RMT
potential $V(M)$.

The purpose of the present letter is two-fold. First, we shall turn the
question around, and ask for the consequences of including higher-order
terms in the RMT potential $V(M)$. Remarkably, there is
a huge degree of universality at work here. To leading order in $1/N^2$
the precise result (\ref{Wesz}) changes, {\em but the leading-order
contribution in $\sqrt{\lambda}$ remains generically
unaffected}\footnote{We shall comment on the precise condition below.}.
Moreover, all higher-order terms in $1/N^2$ are generically
insensitive to the precise form of the potential $V(M)$, but depend
only on a finite number of ``moments'' $M_k$. These statements of
universality follow from a series of results derived in connection with the
RMT approach to 2D quantum gravity \cite{AJM,ACM}. Our second purpose is
to advocate the use of the so-called loop insertion method
\cite{ACKM} to compute these circular Wilson loop
expectation values. We believe this technique is superior to the
orthogonal polynomial method in the present context, independently of
whether one wants to restrict oneself to the Gaussian potential or not.
To illustrate this, we show how to compute the most general $k$-loop
correlation function for any potential $V(M)$, to all orders in the
$1/N^2$ expansion. We first briefly, in
the section below, recall some of the main results from Random Matrix
Theory.

\section{The $k$-point function to all orders in $1/N^2$}

Our starting point is the observation that the
loop expectation value $\langle (1/N)\t \exp[\sqrt{\lambda}M]
\rangle$ is the Laplace transform ${\cal L}$
of the 1-point function or resolvent $G(p)$.
Defining
\beq
G(p) ~\equiv~ \left\langle \frac{1}{N} \t \frac{1}{p - M}\right\rangle
\eeq
we immediately have
\beq
{\cal L}^{-1}[G(p)](x) ~=~
\left\langle \frac{1}{N}\t ~e^{xM} \right\rangle ~\equiv~ W(x)~.
\eeq
Here, the expectation value with respect to the Random Matrix Theory
partition function of the unitary ensemble
\beq
{\cal Z} ~\equiv~ \int\! dM \exp\left[-N \t V(M)\right]\ \ ,\ \ \ \
V(M) ~=~ \sum_{k=1}^{\infty}\frac{1}{k}g_k M^k ~.
\label{V}
\eeq
is defined
with a general potential $V(M)$
in the standard way.

To compute a Wilson loop in an arbitrary representation of $U(N)$
we need also more general expectation values
\beq
\left\langle \frac{1}{N}\t ~e^{x_1M}\cdots \frac{1}{N}\t ~e^{x_kM}
\right\rangle \ .
\eeq
These can be computed from the corresponding connected loop expectation
values
\beq
W(x_1,\ldots,x_k) ~\equiv~
N^{k-2} \left\langle \t ~e^{x_1M}\cdots \t~ e^{x_kM}
\right\rangle_{conn.}
\ =\ \sum_{j=0}^{\infty} \frac{1}{N^{2j}}\ W_j(x_1,\ldots,x_k)
\ .
\eeq
The latter will follow similarly
from the connected $k$-point correlation
function ($k$-point resolvent)
\beq
G(p_1,\ldots,p_k) ~\equiv~ N^{k-2}\left\langle \t \frac{1}{p_1 - M} \cdots
\t \frac{1}{p_k - M}\right\rangle_{conn.}
\ =\ \sum_{j=0}^{\infty} \frac{1}{N^{2j}}\ G_j(p_1,\ldots,p_k)
\label{Gkdef}
\eeq
by $k$ inverse Laplace transforms. For that purpose we will rely on the
very powerful results for the connected $k$-point function which
have been given explicitly in ref.
\cite{ACKM} to all orders in the $1/N^2$ (genus) expansion.
The reader familiar with this material can skip it, and go straight
to section 3.

\subsection{{\em The 1-point function}}

Starting with the 1-point function
the general solution is of the following form
\cite{ACKM}
\beq
G_k(p) ~=~ \sum_{n=1}^{3k-1}\left( A_k^{(n)} \chi^{(n)}(p) +
                                   D_k^{(n)} \psi^{(n)}(p) \right)
\ , \ \ k\geq 1.
\label{Gk-1}
\eeq
The constants $A_k^{(n)}$ and $D_k^{(n)}$ are rational functions of the
``moments''
\be
M_k &\equiv&
\oint_C \frac{d\omega}{2\pi i} \frac{V'(\omega)}{(\omega-a)^{k+1/2}
(\omega-b)^{1/2}} ~=~ g_{k+1} + g_{k+2}\left(\frac12 b + (k+\frac12)a
\right) +\ldots\cr
J_k &\equiv&
\oint_C \frac{d\omega}{2\pi i} \frac{V'(\omega)}{(\omega-a)^{1/2}
(\omega-b)^{k+1/2}}~=~ g_{k+1} + g_{k+2}\left(\frac12 a + (k+\frac12)b\right)
+\ldots
\label{Mk}
\ee
and depend on at most $2(3k-1)$ of them. These moments encode in a universal
way the dependence on the infinite set of coupling constants
$\{g_k\}$, and we
will give a few examples below. The functions $\chi^{(n)}(p)$ and
$\psi^{(n)}(p)$ are the basis functions needed for the solution of the loop
equation. They are given explicitly in ref. \cite{ACKM}.
All we need to know here is that $\chi^{(n)}(p)$ and
$\psi^{(n)}(p)$ are linear combinations of the functions
$\phi_a^{(k)}(p)=(p-a)^{-k-1/2}(p-b)^{-1/2}$ and
$\phi_b^{(k)}(p)=(p-b)^{-k-1/2}(p-a)^{-1/2}$, respectively,
up to order $k=n$. They also depend on the moments $M_k$ and $J_k$ up to
$k=n$. Since the inverse Laplace transforms of $\phi_a^{(k)}(p)$
and $\phi_b^{(k)}(p)$ are known in closed form, we immediately read off
the corresponding expansion of the Wilson loop $W(x)$.
For a Gaussian
potential $V(M)=\frac12 g_2 M^2$ we have $b=-a$ and $M_1=J_1=4/a^2=g_2$.
All higher moments $M_{k\geq2}$ and $J_{k\geq2}$ vanish.
In particular the correlators then contain
less terms and thus differ from  the general expression.

The result given in eq. (\ref{Gk-1}) gives only the $1/N^{2k}$
corrections to the 1-point function $G(p)$ for $k\geq 1$.
The leading order result of $k=0$
is non-universal and depends explicitly on all the coupling constants
in the potential $V(M)$. It can be written \cite{Mak}
\beq
G_0(p) ~=~ \frac12
\oint_C \frac{d\omega}{2\pi i} \frac{V'(\omega)}{p-\omega}
\sqrt{\frac{(\omega-b)(\omega-a)}{(p-b)(p-a)}} \ +\
\frac{1}{\sqrt{(p-b)(p-a)}} ~.
\label{G0-1}
\eeq
The boundary conditions that fix the end-points of the cut $[b,a]$
are
\beq
\delta_{1,k} ~=~ \frac12 \oint_C \frac{d\omega}{2\pi i}\omega^k V'(\omega)
\frac{1}{\sqrt{(\omega-b)(\omega-a)}} \ ,\ \ k=0,1 
\label{bc}
\eeq
which follow from the requirement that
$G_0(p)\sim \frac{1}{p}$ for $p\to\infty$.
In order to illustrate the non-universality let us give an explicit example for
a symmetric sixth-order potential:
\beq
G_0(p) ~=~ \frac12 \left[ g_2p + g_4p^3 + g_6p^5 -
\left( g_2 + \frac12 g_4a^2 + \frac38 g_6a^4 + (g_4 + \frac12 g_6a^2)p^2
+ g_6p^4\right)\sqrt{p^2-a^2} \right] \ .
\label{G0-1sext}
\eeq
The endpoint $a$ of the cut (which is now symmetric) $[-a,a]$ is given by
eq. (\ref{bc}):
\beq
1~=~ \frac14 g_2a^2 + \frac{3}{16} g_4a^4 + \frac{5}{32}g_6a^6 \ .
\label{defb}
\eeq

\subsection{{\em Higher k-point functions}}

Higher $k$-point functions can be obtained
by applying successively the
loop insertion operator \cite{ACKM}
\beq
\frac{d}{dV(p)} ~\equiv~ - \sum_{k=1}^{\infty}\frac{k}{p^{k+1}}\frac{d}
{dg_{k}} ~,
\eeq
starting from the resolvent itself:
\beq
G(p_1,\ldots,p_k) ~=~ \frac{d}{dV(p_{k})}\ldots\frac{d}{dV(p_{2})}G(p_{1}) ~.
\eeq
The planar 2-point function obtained in this way from eq. (\ref{G0-1}) is
universal, the well-known result of Ambj{\o}rn, Jurkiewicz and
Makeenko \cite{AJM}:
\beq
G_0(p,q) ~=~ \frac{1}{4(p-q)^2}\left[
\frac{(p-b)(q-a)+(p-a)(q-b)}{\sqrt{(p-b)(p-a)(q-b)(q-a)}} ~-~ 2
\right] \ .
\label{G0-2}
\eeq
As a consequence all $k$-point functions for $k\geq 2$ are universal to all
orders in $1/N^2$, because the derivatives $d/dV (p)$ acting
on the universal variables $b,a,M_k,J_k$ can be expressed again in terms of
these variables and the functions $\phi_b^{(k)}(p)$
and $\phi_a^{(k)}(p)$. They are again given explicitly in \cite{ACKM}.

Apart from $G_0(p,q)$
all universal $k$-point functions $G_j(p_1,\ldots,p_{k})$
factorize with respect to their arguments
into a finite number of terms to all orders in $1/N^2$.
This follows from the fact that
once a function $G_j(p_1,\ldots,p_{k-1})$ is factorized, the application of
the loop insertion operator $d/dV (p_k)$ only adds factors which are
functions of $p_k$. At genus zero the 3-point function factorizes
\be
G_0(p,q,r) &=&\frac{a-b}{8}\left[\frac{1}{M_1}
\frac{1}{(p-a)^{3/2}(p-b)^{1/2}}\ \frac{1}{(q-a)^{3/2}(q-b)^{1/2}}\
\frac{1}{(r-a)^{3/2}(r-b)^{1/2}} \right. \cr
&&\ \ \ \ \ \ -\ \left.\frac{1}{J_1}
\frac{1}{(p-a)^{1/2}(p-b)^{3/2}}\ \frac{1}{(q-a)^{1/2}(q-b)^{3/2}}\
\frac{1}{(r-a)^{1/2}(r-b)^{3/2}}
\right]\ .
\ee
By applying successively the loop insertion operator one sees that
all $G_0(p_1,\ldots,p_k)$ also factorize for $k>3$.
At higher genus factorization follows immediately
from the 1-point function eq. (\ref{Gk-1}) (see the example for genus one
below).
This factorization property makes the
successive application of $k$ inverse Laplace transformations particularly
simple. The only exception is $G_0(p,q)$, which requires a little more
work.

Let us finally mention that the same procedure we have described here applies
immediately to the complex matrix model \cite{Mor} and to the so-called
reduced hermitian matrix model. There, a closed expression
has been derived for all planar $k$-point correlators with $k\geq 2$
\cite{AJM,AHW}. All higher genus corrections in $1/N^2$ are also available
for both the complex \cite{AKM} and reduced hermitian models \cite{A}.
Precisely as for the hermitian case treated below, we can simply read off
the corresponding loop correlation functions from the inverse Laplace
transformations.

\section{Wilson loops from Laplace transforms}

We are now ready to compute Wilson loop expectation values by means
of inverse Laplace transforms. We begin with the Wilson loop in the
fundamental representation of $U(N)$. As already mentioned, this expectation
value as computed in Random Matrix Theory is not universal. Consider
as an example the sixth-order (symmetric) potential from
eq. (\ref{G0-1sext}). To leading order in $1/N^2$ we get
\beq
W_0(x) ~=~ \frac{2}{ax}I_1(ax) \ +\ \frac38\left( g_4+\frac54 g_6 a^2\right)
\frac{a^3}{x}I_3(ax) \ +\ \frac{5}{32}g_6\ \frac{a^5}{x}I_5(ax) ~.
\label{W0-1sextic}
\eeq
This shows that the natural variable is the combination $ax$, and we
identify $a \equiv \sqrt{\lambda}$. The asymptotic behavior for large
$\lambda$ appears to depend in a complicated way on the coupling constants
$g_4$ and $g_6$. But from the constraint (\ref{defb}) it follows that
in order to achieve $a \to \infty$ 
we require at least\footnote{For an arbitrary
symmetric potential the constraint (\ref{bc}) can be written
$1=\frac12\sum_{k=0}^\infty g_{2k} {{2k} \choose {k}} \frac{a^{2k}}{4^k}$ .}
$g_{2n} \sim a^{-2n}$ (or even higher suppression in 1/a). 
Because the leading term in the asymptotic
Bessel-function expansion at large argument 
is $I_n(z) \sim \exp[z]/\sqrt{2\pi z}$ for all fixed $n$, the large-$\lambda$
behavior is therefore
\beq
W_0(x) \sim const.~\lambda^{-3/4}\exp[x\sqrt{\lambda}] ~,
\eeq
in agreement with the Gaussian result \cite{ESZ,DG} when we set $x=1$.
So the {\em leading}
large-$\lambda$ behavior is unaffected by higher order terms in the
potential $V(M)$.

We now turn to the first $1/N^2$ correction to this result. For simplicity,
let us here again restrict ourselves to symmetric potentials (but see the
Appendix for the most general expressions).
The universal function $W_1$ follows from
the genus-one 1-point function $G_1(p)$ for a symmetric potential
\beq
G_1(p) ~=~ \frac{a^2}{4M_1} \ \frac{1}{(p^2-a^2)^{5/2}} \ -\
\frac{aM_2}{8(M_1)^2}  \ \frac{1}{(p^2-a^2)^{3/2}}
\label{G1-1S} ~.
\eeq
Taking the inverse Laplace transform, we get the general result
\beq
W_1(x) ~=~ \frac{1}{12M_1}\ x^2 I_2(ax) ~-~\frac{M_2}{8(M_1)^2}\ xI_1(ax) ~.
\label{W1-1S}
\eeq
Specifically, for the same symmetric sixth-order potential as above we have
\beq
M_1 ~=~ g_2 \ +\ \frac32 a^2 g_4 \ +\ \frac{15}{8}a^4 g_6 \ , \ \ \ \
M_2 ~=~ 2 a g_4 \ +\ 5 a^3 g_6 \ \ \ .
\label{Mexplicit}
\eeq
For a Gaussian potential this reproduces the result of ref.
\cite{DG}: $W_1(\lambda) = \lambda I_2(\sqrt{\lambda})/48$. Again we find that
the leading large-$\lambda$ behavior
is unaffected by higher order terms in the
potential, after using $M_1 \sim a^{-2}, M_2 \sim a^{-3}$ as $a \to \infty$:
$W_1(\lambda) \sim const. \lambda^{3/4}\exp[\sqrt{\lambda}]$. Note that
the second term in (\ref{W1-1S}) is subleading in this limit. We have also
computed the general genus-two ($1/N^4$) contribution $G_2(p)$,
and confirmed the result
of ref. \cite{DG} for the special case of a Gaussian potential. Clearly, the
leading $\lambda$-behavior is also here of the same form for any
generic potential.

We next turn to higher $k$-point correlation functions of these fundamental
Wilson loops, needed for the evaluation of Wilson loops in arbitrary
representations of $U(N)$. The universal 2-loop result (\ref{G0-2}) is seen
to not factorize in $p$ and $q$. One may use the convolution theorem to
evaluate the needed double inverse Laplace transform, but it is simpler to
note that
\beq
\frac{\partial}{\partial a^2} G_0(p,q) ~=~ \frac14
\frac{pq + a^2}{(p^2-a^2)^{3/2}(q^2-a^2)^{3/2}}
\eeq
{\em does} factorize. It is thus elementary to get the inverse Laplace
transforms,
\beq
\frac{\partial}{\partial a^2} W_0(x,y) ~=~ \frac{xy}{4}
\left[ I_0(ax)I_0(ay)\ +\  I_1(ax)I_1(ay) \right]
\eeq
which we integrate up to
\beq
W_0(x,y) ~=~  \frac{xy}{2} \int_0^a \! du\ u
\left[ I_0(ux)I_0(uy)\ +\  I_1(ux)I_1(uy) \right] ~,
\label{W0-2int}
\eeq
after fixing the integration constant by comparing \cite{AJM}
\beq
G_0(p,p) ~=~ \frac{a^2}{4(p^2-a^2)^2} ~=~ \frac{1}{p^4}\ \frac{a^2}{4}\ +\
{\cal O}\left(\frac{1}{p^6}\right)
~=~ \frac{1}{p^4}\langle \t M \t M\rangle_{conn.}\ +\
{\cal O}\left(\frac{1}{p^6}\right)
\eeq
and
\beq
W_0(x,y) ~=~ xy\ \langle \t M \t M\rangle_{conn.} \ +\
{\cal O}\left(x^2y^2, xy^3, x^3y\right) \ .
\eeq
Performing the integral in eq. (\ref{W0-2int}) we get
for an arbitrary symmetric potential
\beq
W_0(x,y) ~=~ \frac{axy}{2(x+y)}\left[ I_0(ay)I_1(ax)\ +\
I_0(ax)I_1(ay)\right] ~.
\label{W0-2S}
\eeq
As a check on this result, we note that it vanishes when one or both of the
arguments $x$ and $y$ become equal to zero. 
Finally, at equal arguments we have
\beq
W_0(x,x) ~=~ \frac{ax}{2}I_0(ax)I_1(ax) ~.
\label{W0-2Sxx}
\eeq
As an example, let us evaluate $N^{-2}\langle (\t \exp[M])^2\rangle$, which in
ref. \cite{DG} is denoted by $W_{1,1}$. We get
\beq
W_{1,1} ~=~ (W_0(1))^2 + \frac{1}{N^2}[W_0(1,1) + 2W_0(1)W_1(1)]
+ {\cal O}\left(\frac{1}{N^4}\right)~,
\eeq
where all functions on the right hand side have been given above. In
particular, we find
\beq
W_{1,1} ~=~ \frac{4}{\lambda}I_1(\sqrt{\lambda})^2 
+ \frac{\sqrt{\lambda}}{2N^{2}}
\left[I_0(\sqrt{\lambda})I_1(\sqrt{\lambda}) +
\frac{1}{6}I_1(\sqrt{\lambda})I_2(\sqrt{\lambda})\right]
+ {\cal O}\left(\frac{1}{N^4}\right)
\eeq
in the Gaussian case. The leading large-$\lambda$ behavior of the
$1/N^2$ correction goes like $\exp[2\sqrt{\lambda}]$,
which is now ready to be compared with string theory through the
AdS/CFT correspondence.

It is straightforward to go on to arbitrarily high order in $1/N^2$ for
any $n$-point correlation function. As a last example, consider the
universal 3-point function $W_0(p,q,r)$ for an arbitrary symmetric potential.
We find:
\be
W_0(x,y,z) &=& xyz \frac{a}{2M_1}\left[ I_1(ax)I_0(ay)I_0(az)
\ +\ I_0(ax)I_1(ay)I_0(az)
\right. \cr
&&\ \ \ \ \ \ \ \ \ \ \left. + I_0(ax)I_0(ay)I_1(az)\ +\ I_1(ax)I_1(ay)I_1(az)
\right]
\label{W0-3S}
\ee
where for the Gaussian case $M_1 = g_2 = 4/a^2 = 4/\lambda$.
The most general result for the genus expansion of the resolvent
can be written
\beq
W_k(x) ~=~ \sum_{n=1}^{3k-1}\left( A_k^{(n)} {\cal L}^{-1}[\chi^{(n)}(p)](x) +
                                   D_k^{(n)} {\cal L}^{-1}[\psi^{(n)}(p)](x)
\right)
\ , \ \ k\geq 1.
\label{wk-1}
\eeq
where the inverse Laplace transforms are known explicitly. Any $n$-point
function for $n\geq 2$ can then be found to all orders in the $1/N^2$
expansion by the iteration procedure described above. We have thus succeeded
in determining all $n$-point Wilson loop correlation functions to all
orders in the $1/N^2$ expansion. They are all given by appropriate products
of $n$ modified Bessel functions.

\section{Other universality classes}

By fine-tuning the RMT potential one can reach new (multi-critical)
universality classes that would invalidate the above conclusions.
To see what happens in such cases, it suffices
to focus on the 1-point function, the Wilson loop itself, in the planar
limit. First note that, with $\rho(\zeta)$ indicating the eigenvalue density
\beq
\left\langle \frac{1}{N} \t ~e^{xM} \right\rangle ~=~
\frac{1}{\cal Z}\int\! dM~ \frac{1}{N} \t~e^{xM}~
\exp\left[-N\t V(M)\right] 
~=~\int_{-1}^{1}\! d\zeta_1~ \exp(ax\zeta_1)~\rho(\zeta_1)
\label{generalW}
\eeq
where for simplicity we have restricted ourselves to even potentials
$V(M)$, and the cut has been rescaled to lie on the interval [-1,1].
By inserting the Wigner semi-circle law $\rho(\zeta)
= (2/\pi)(1- \zeta^2)^{1/2}$ corresponding
to a Gaussian potential we of course just recover the result (\ref{Wesz})
with $a=\sqrt{\lambda}$.
(The form (\ref{generalW}) also immediately shows the non-universality of
this result, away from the large-$\lambda$ limit). Multicritical densities
belonging to multicritical universality classes near the soft edge
can be chosen
\beq
\rho_{m}(\zeta) ~\sim~ (1 - \zeta^2)^{m+1/2} ~.
\eeq
According to eq. (\ref{generalW}) the corresponding $m$th multicritical
Wilson loop is
\beq
W_0(x) ~=~
\frac{(k+1)!2^{k+1}}{(ax)^{k+1}}I_{k+1}(ax)~\sim~ 
const.
\frac{e^{x\sqrt{\lambda}}}{{\lambda}^{(2k+3)/4}} ~.
\label{multi}
\eeq
The leading large-$\lambda$ behavior thus differs in the prefactor from
that of generic non-critical potentials, although the exponential
form $\sim \exp[x\sqrt{\lambda}]$ is retained. We have explicitly checked
eq. (\ref{multi}) for $m=1,2$ by tuning the potential accordingly
in eq. (\ref{G0-1sext}). For this special choice of coupling constants
the Bessel functions in eq. (\ref{W0-1sextic}) precisely combine to give eq.
(\ref{multi}).

It is tempting to speculate that the universality classes of
multicritical points could correspond to circular Wilson loops of
different gauge theories at conformal points. But we have no evidence
to support such a claim.

\noi\noindent
{\bf Acknowledgments}\\
We thank R.D. Ball and J. Plefka for helpful discussions.
This work was supported in part by EU TMR grant no.
ERBFMRXCT97-0122.

\appendix

\section{Appendix}

In this appendix we give, for completeness, a few examples to illustrate
that the method we have presented
also works for an arbitrary potential that is not restricted
to be even. We start with a closed expression for the
non-universal $W_0(p)$ which can be derived from eq. (\ref{G0-1})
\be
W_0(x) &=& \int_0^x \!du\ \exp\left(u\frac{a+b}{2}\right)
I_0\left(u\frac{a-b}{2}\right)\frac12
\oint_C \frac{d\omega}{2\pi i}\  V'(\omega)
\sqrt{\frac{(\omega-b)(\omega-a)}{(p-b)(p-a)}}\exp[\omega(x-u)]
\cr
&+& \exp\left(x\frac{a+b}{2}\right)I_0\left(x\frac{a-b}{2}\right)
\label{W0-1}
\ee
Next, we find the $1/N^2$ contribution:
\be
W_1(x) &=& x^2\frac{1}{16}\exp\left(x\frac{a+b}{2}\right)
\left\{
\frac{1}{M_1}\left[\frac12 I_0\left(xc\right) +
\frac23 I_1\left(xc\right) +
\frac16 I_2\left(xc\right)\right] \right. \cr
&&\ \ \ \ \ \ \ \ \ \ \ \ \ \ \ \ \ \ \ \ \ \ \ \ \  +\  \left.
\frac{1}{J_1}\left[ \frac12 I_0\left(xc\right)
- \frac23 I_1\left(xc\right) +
\frac16 I_2\left(xc\right)\right]
\right\}  \cr
&-&  x\frac{1}{16}\exp\left(x\frac{a+b}{2}\right)\left\{
\left( \frac{M_2}{(M_1)^2} + \frac{1}{cM_1}\right)
\left[ I_0\left(xc\right)
+ I_1\left(xc\right)\right] \right. \cr
&&\ \ \ \ \ \ \ \ \ \ \ \ \ \ \ \ \ \ \ \ \ \ \ \ \ +\  \left.
\left( \frac{J_2}{(J_1)^2} - \frac{1}{cJ_1}\right)
\left[ I_0\left(xc\right)
- I_1\left(xc\right)\right]
\right\} \ ,
\label{W1-1}
\ee
where we have defined $c\equiv (a-b)/2$.
Note the appearance of the exponential prefactor
$\exp[x(a+b)/2]$ which disappears for the symmetric potential with $b=-a$.
The result for a generic symmetric potential eq. (\ref{W1-1S})
is easily recovered by setting $J_k=(-1)^{k+1}M_k$ and $c=a$.
As a last example let us also give the planar connected 3-loop function
\be
W_0(x,y,z) &=& xyz\frac{a-b}{8}\exp\left(xyz\frac{a+b}{2}\right)
\mbox{\Big\{} \frac{1}{M_1}
\left[I_0\left(xc\right)+I_1\left(xc\right)\right]
\left[I_0\left(yc\right)+I_1\left(yc\right)\right]
\left[I_0\left(zc\right)+I_1\left(zc\right)\right]
\cr
&& \ \ \ \ \ \ \ \ \ \ \ \ \ \ \ \ \ \ \ \ \ \ \ \ \ \ \ \ \ \ \ \ \
-\ \frac{1}{J_1}
\left[I_0\left(xc\right)-I_1\left(xc\right)\right]
\left[I_0\left(yc\right)-I_1\left(yc\right)\right]
\left[I_0\left(zc\right)-I_1\left(zc\right)\right]
\mbox{\Big\}}. \cr
&&
\label{W0-3}
\ee
Also the general expression for the 2-loop function can be evaluated
for general asymmetric potentials, using a variant of the method
described above for the symmetric case. Proceeding iteratively, this
again determines all $n$-point functions to all orders in the
$1/N^2$-expansion, now for potentials that are not necessarily
symmetric.


\begin{thebibliography}{X}

\bibitem{M}J. Maldacena, Adv. Theor. Math. Phys. {\bf 2} (1998) 231.
\bibitem{RY}S. Rey and J. Yee, hep-th/9803001.
\bibitem{M1}J. Maldacena, Phys. Rev. Lett. {\bf 80} (1998) 4859.
\bibitem{ESZ}J.K. Erickson, G.W. Semenoff and K. Zarembo,
Nucl. Phys. {\bf B582} (2000) 155.
\bibitem{BCFM}D. Berenstein, R. Corrado, W. Fischler and J. Maldacena,
Phys. Rev. {\bf D59} (1999) 105023.
\bibitem{DGO}N. Drukker, D.J. Gross and H. Ooguri, Phys. Rev. {\bf
D60} (1999) 125006.
\bibitem{DG}N. Drukker and D.J. Gross, hep-th/0010274.
\bibitem{G}D.J. Gross, talk at the Strings 2001 meeting.
\bibitem{AJM}J. Ambj{\o}rn, J. Jurkiewicz and Yu. Makeenko,
Phys. Lett. {\bf B251} (1990) 517.
\bibitem{ACM}J. Ambj{\o}rn, L. Chekhov and Yu. Makeenko, Phys. Lett.
{\bf B282} (1992) 341.
\bibitem{ACKM}J. Ambj{\o}rn, L. Chekhov, C.F. Kristjansen and
Yu. Makeenko, Nucl. Phys. {\bf B404} (1993) 127.
\bibitem{Mak}Yu. Makeenko, Mod. Phys. Lett. {\bf A6} (1991) 1901.
\bibitem{Mor} T.R. Morris,  Nucl. Phys. {\bf B356} (1991) 703.
\bibitem{AHW}J. Ambj{\o}rn, M. Harris, and M. Weis, Nucl. Phys. {\bf B504}
(1997) 482.
\bibitem{AKM}J. Ambj{\o}rn, C.F. Kristjansen and Yu. Makeenko,
 Mod. Phys. Lett. {\bf A7} (1992) 3187.
\bibitem{A}G. Akemann, Habilitation Thesis, Ruprecht-Karls-Universit\"at
Heidelberg, May 2000, Preprint MPIH-V11-2000.


\end{thebibliography}
\end{document}